\documentclass[prd,
twocolumn,
superscriptaddress,nofootinbib,%
tightenlines,shownopacs]{revtex4}
\usepackage{mathrsfs}
\usepackage{latexsym,bm}
\usepackage{CJK}
\usepackage{graphicx}
\usepackage{indentfirst}
\usepackage{slashed}
\usepackage{amsmath}
\usepackage{amssymb}
\usepackage{color}
\usepackage{hyperref}
\usepackage{epsfig}
\hypersetup{CJKbookmarks=true}
\usepackage[titletoc]{appendix}
\usepackage{tabularx}
\usepackage{lipsum}

\newcommand{\be}{\begin{equation}} \newcommand{\ee}{\end{equation}}
\newcommand{\ba}{\begin{array}{c}} \newcommand{\ea}{\end{array}}
\newcommand{\bea}{\begin{eqnarray}} \newcommand{\eea}{\end{eqnarray}}

\newcommand{\no}{\nonumber}

\newcommand{\mpi}{M_\pi}
\newcommand{\mn}{m_N}
\newcommand{\md}{m_\Delta}
\newcommand{\gA}{\mathring{g}_A}
\newcommand{\go}{g_1}

\newcommand{\D}{\displaystyle}

\allowdisplaybreaks

\begin{document}
\title{\Large Extraction of nucleon axial charge and radius from lattice QCD results using baryon chiral perturbation theory}
\author{De-Liang~Yao}
\email{deliang.yao@ific.uv.es}
\author{Luis~Alvarez-Ruso}
\email{luis.alvarez@ific.uv.es}
\author{Manuel~J.~Vicente-Vacas}
\email{vicente@ific.uv.es}
\affiliation{Departamento de F\'isica Te\'orica and Instituto de F\'\i sica Corpuscular (IFIC), Centro Mixto UVEG-CSIC, Valencia, Spain}

\begin{abstract}
We calculate the nucleon axial form factor up to the leading one-loop order in a covariant chiral effective field theory with the $\Delta(1232)$ resonance as an explicit degree of freedom.  We fit the axial form factor to the latest lattice QCD data and pin down the relevant low-energy constants. The lattice QCD data, for various pion masses below $400$~MeV, can be well described up to a momentum transfer of $\sim 0.6$~GeV. 
The $\Delta(1232)$ loops contribute significantly to this agreement. Furthermore, we extract the axial charge and radius based on the fitted values of the low energy constants. The results are: $g_A=1.237(74)$ and $\langle r_A^2\rangle =0.263(38)~{\rm fm}^2$. The obtained coupling $g_A$ is consistent with the experimental value if the uncertainty is taken into account. The axial radius is below but in agreement with the recent extraction from neutrino quasi-elastic scattering data on deuterium, which has large error bars. Up to our current working accuracy, $r_A$ is predicted only at leading order, i.e., one-loop level. A more precise determination might need terms of $\mathcal{O}(p^5)$. 
\end{abstract}
\maketitle

\section{Introduction}
Nucleon form factors are basic quantities characterizing the nucleon and hence are of fundamental importance in our understanding of hadron structure.  One of the most important quantities is the nucleon isovector axial form factor. This is not only because its value at zero momentum transfer defines the prominent nucleon axial charge $g_A$, but also due to the relevance of its momentum-transfer dependence to experimental processes such as (quasi)elastic neutrino-nucleon scattering, whose precise understanding is crucial to achieve the precision goals in the determination of neutrino-oscillation parameters~\cite{Alvarez-Ruso:2017oui}.

The nucleon axial charge describes the difference of the spins carried by the $u$ and $d$ quarks in the proton. Experimentally, its value is very accurately determined through neutron $\beta$-decay, $g_A=1.2723(23)$~\cite{Patrignani:2016xqp}. Extractions of the momentum-transfer dependence of the axial form factor are far more demanding and rely on pion electroproduction, neutrino-induced charged-current quasielastic scattering on deuteron targets and weak capture in muonic hydrogen ~\cite{Liesenfeld:1999mv,Bernard:2001rs,Bodek:2007ym,Meyer:2016oeg,Hill:2017wgb}. The determination by the MiniBooNE neutrino experiment~\cite{AguilarArevalo:2010zc} should not be used as a reference because data were taken on a $^{12}$C target and included a sizable contribution from two-nucleon currents; details can be found for instance in Sec.~III of Ref.~\cite{Katori:2016yel} and references therein.  
 
 On the side of lattice QCD, there are several determinations of the axial charge, for instance in Refs~\cite{Yamazaki:2008py,Capitani:2012gj,Horsley:2013ayv,Bhattacharya:2016zcn,Yoon:2016jzj,Liang:2016fgy}. Yet,  values lower than the experimental one have been recurrently obtained.  Recently, progress has been made by using improved algorithms, and experimentally consistent results have been obtained~\cite{Berkowitz:2017gql,Alexandrou:2017hac,Capitani:2017qpc}. More importantly, lattice studies of the momentum-transfer dependence of the axial form factor have also been accumulated~\cite{Yamazaki:2009zq,Bratt:2010jn,Alexandrou:2010hf,Green:2017keo,Alexandrou:2017hac,Capitani:2017qpc,Rajan:2017lxk}. On the one hand, these results have been analyzed using the dipole ansatz [Eq.~(\ref{eq:dipole}) below] and, more recently, with the more general $z$-expansion~\cite{Bhattacharya:2011ah}. On the other hand, the extrapolation of the lattice results to the physical limit has been performed with formulae obtained in chiral perturbation theory (ChPT)~\cite{Weinberg:1978kz,Gasser:1983yg,Gasser:1984gg,Bernard:1995dp}. However, instead of the above procedure, it is more efficient and systematic to study both the pion mass and momentum-transfer dependences of the nucleon axial form factor in baryon chiral perturbation theory (BChPT).

ChPT has proved to be a powerful tool at low-energies and is widely used in modern hadronic physics. The axial form factor up to one-loop level has been calculated in Refs.~\cite{Bernard:1993bq,Fearing:1997dp,Bernard:1998gv}, with the framework of heavy-baryon ChPT (HBChPT), and in Refs.~\cite{Schindler:2006it,Ando:2006xy}, within relativistic baryon ChPT, using different renormalization schemes such as the infrared regularization (IR) prescription~\cite{Ellis:1997kc,Becher:1999he} and the extended-on-mass-shell (EOMS) scheme~\cite{Gegelia:1999gf,Fuchs:2003qc}. Therein, the explicit contribution involving the $\Delta(1232)$-resonance is included only in Ref.~\cite{Bernard:1998gv} using HBChPT. The non-analytical part of the axial form factor at two-loop level can be found in Ref.~\cite{Bernard:1996cc}; this study has been recently summarized in Ref.~\cite{Krebs:2016rqz}. As for $g_A$, calculations are more advanced: a full two-loop result of HBChPT is available in Ref.~\cite{Bernard:2006te} and a full one-loop result with EOMS scheme in covariant BChPT was obtained in Ref.~\cite{Chen:2012nx}. To the best of our knowledge, only  the non-relativistic chiral results of $g_A$ have been compared to lattice QCD data~\cite{Hemmert:2003cb,AliKhan:2003ack,Procura:2006gq}.  However, due to the progress of lattice QCD calculations, it is now feasible to compare the chiral axial form factor to  the wealth of lattice QCD data. 
The octet-baryon axial charges have also been extensively studied using HBChPT~\cite{Jenkins:1991es,FloresMendieta:2012dn,CalleCordon:2012xz}, IR~\cite{Zhu:2000zf} and EOMS~\cite{Ledwig:2014rfa}. The approach of Ref.~\cite{Ledwig:2014rfa} closely resembles the one adopted here.

In the present work, we calculate the nucleon axial form factor up the leading one-loop order in a covariant BChPT with explicit $\Delta(1232)$ (simply $\Delta$ from now on). By applying the EOMS scheme, the ultraviolet (UV) divergences and polynomials of power-counting breaking (PCB) terms from the loops are absorbed in the parameters appearing in the chiral effective Lagrangian. With these choices, our calculation is encompassed in a unified framework, which can be  applied to all possible observables. In such a consistent treatment, low energy constants (LECs) extracted in specific processes can be reliably used in other studies. In particular, we adopt $\Delta$ LECs obtained earlier from $\pi N$ scattering within the same approach and at the same chiral order.

Our covariant chiral representation of the axial form factor has the correct power counting and keeps the proper analytical properties, being hence appropriate for performing chiral extrapolations. We fit our chiral result of the axial form factor to the latest lattice QCD data~\cite{Alexandrou:2017hac,Capitani:2017qpc,Rajan:2017lxk} at various pion masses and up to $Q^2\equiv-t=0.36$~GeV$^2$ with $t$ being the momentum-transfer squared. For comparison, the case without the contribution of explicit $\Delta$ is investigated as well. The fitted data are well described and the involved low-energy constants are pinned down. Based on the fitted values of the LECs, we extract the axial charge and radius. Their expressions are shown explicitly for easy reference in the future.
 
 The paper is organized as follows. Section~\ref{sec:AxialFF} contains the details of our calculation within BChPT. The definitions of axial form factor, charge and radius are introduced in section~\ref{sec:definitions}. The effective Lagrangians and the chiral results are specified in sections~\ref{sec:Lag} and~\ref{sec:loop}, respectively. The numerical study is described in section~\ref{sec:lat}. Our fitting procedure is explained in section~\ref{sec:fit}. In section~\ref{sec:gara}, the extractions of the axial charge and radius are discussed. We summarize in section~\ref{sec:sum}. Explicit loop expressions of the axial charge and radius are relegated to Appendix~\ref{sec:ologa}.

\section{Axial form factor in BChPT\label{sec:AxialFF}}
\subsection{Definitions\label{sec:definitions}}
The isovector axial current of light quarks in QCD is written as the local bilinear operator
\bea
A_\mu^i(x)=\bar{q}(x) \gamma_\mu\gamma_5\frac{\tau^i}{2} q(x)  
\eea
of the quark-field doublet $q=(u,d)^T$; $\tau^i$ ($i=1-3$) are the Pauli isospin matrices. In the isospin limit, the transition matrix element of the this current between nucleon states is decomposed as
\bea
&\langle N(p\prime)|A_\mu^i(0)| N(p)\rangle =&  \nonumber \\
&\bar{u}(p\prime) \bigg[\D \gamma_\mu G_A(t)+\frac{\D q_\mu}{ 2m_N}G_{P}(t)\bigg]\gamma_5 \frac{\tau^a}{\D 2} u(p)& , 
\eea
where $t\equiv -Q^2 \equiv q^2$ and $q =p\prime-p$ is the momentum transfer; $u(p)$ is the Dirac spinors of a nucleon with momentum $p$, while $m_N$ denotes the nucleon mass. Here $G_A(t)$ and $G_p(t)$ are called the nucleon axial and induced pseudoscalar form factors, respectively. Of our current interest is the axial form factor, which can be expanded in the region of small $t$ as
\bea
G_A(t)=g_A\bigg[1+\frac{1}{6}\langle r_A^2\rangle t+\mathcal{O}(t^2)\bigg]\ ,
\eea
with $g_A \equiv G_A(0)$ being the axial-vector charge. The slope of $G_A(t)$ at $t=0$ defines the nucleon axial radius squared 
\bea\label{eq:rAdef}
\langle r_A^2\rangle=6\frac{\rm d}{{\rm d}t}\bigg[\frac{G_A(t)}{G_A(0)}\bigg]_{t=0} = \frac{12}{M_A^2}\ ,
\eea
related to the so-called axial mass $M_A$ present in the dipole ansatz of the axial form factor
\bea
\label{eq:dipole}
G_A(t) = \frac{G_A(0)}{(1-t/M_A^2)^2}\ ,
\eea
which is extensively used to fit experimental data and lattice QCD results.

\subsection{Chiral effective Lagrangian\label{sec:Lag}}
For the calculation of the axial form factor, we employ the relativistic baryon chiral perturbation theory with pions, nucleons and $\Delta$ as explicit degrees of freedom.  The standard power counting~\cite{Weinberg:1991um} is used for diagrams involving only pion and nucleon lines. For diagrams with $\Delta$ lines, the power counting introduced in 
Refs.~\cite{Hemmert:1996xg,Hemmert:1997ye} and usually referred to as the small-scale expansion (SSE) is applied. In SSE, the mass difference $\delta = m_\Delta-m_N$ is counted as of order ${\cal O}(p)$. Although we adopt SSE for the power counting, in our covariant calculations no expansion in powers of $\delta$ is performed. A different counting rule proposed in Ref.~\cite{Pascalutsa:2002pi} assumes $\delta \sim p^{1/2}$ to preserve the hierarchy $p \sim M_\pi \ll \delta$. As this condition does not hold for many of the lattice results used in the present study, we stick to SSE.

Up to and including ${\cal O}(p^3)$, the following terms of the chiral effective Lagrangian are needed,
\bea
\mathcal{L}_{\rm eff}=\mathcal{L}_{\pi N}^{(1)}+\mathcal{L}_{\pi \Delta}^{(1)}+\mathcal{L}_{\pi N \Delta}^{(1)}+\mathcal{L}_{\pi N}^{(3)}\ ,
\eea
where superscripts and subscripts denote the chiral order and the involved degrees of freedom, respectively. The leading $\pi N$ Lagrangian reads
\bea
{\cal L}_{\pi N}^{(1)}&=&\bar{\Psi}\big\{i\slashed{D}-m
+\frac{1}{2}g \,u^\mu\gamma_\mu\gamma^5\big\}\Psi\ ,
\eea
where $\Psi$ is the nucleon field, while $m$ and $g$ are the bare nucleon mass and the axial charge in the chiral limit, respectively. Discarding external vector fields, the covariant derivative $D_\mu$ acting on the nucleon field and the chiral vielbein $u_\mu$ are defined as
 \bea
 D_\mu&=&\partial_\mu+\Gamma_\mu\ ,\nonumber\\
  \Gamma_\mu&=&\frac{1}{2}[u^\dagger(\partial_\mu-i a_\mu)u+u(\partial_\mu+i a_\mu)u^\dagger]\ ,\nonumber\\
   u_\mu&=&i[u^\dagger(\partial_\mu-i a_\mu)u-u(\partial_\mu+i a_\mu)u^\dagger]\ ,
 \eea 
 and $a_\mu=a_\mu^i\tau^i/2$ is the external axial field. 
 
 After fixing redundant terms~\cite{Tang:1996sq,Krebs:2009bf}, the leading $\pi\Delta$ and $\pi N\Delta$ terms in the Lagrangian can be cast as
\bea\label{eq:LagpiD}
{\cal L}^{(1)}_{\pi\Delta}&=& -\bar{\Psi}_{\mu}^i\xi^{\frac{3}{2}}_{ij} \bigg\{i \left( \slashed{D}^{jk} g^{\mu\nu} + \gamma^\mu\slashed{D}^{jk}\gamma^\nu -\gamma^\mu D^{\nu,jk} \right. \nonumber \\
 &&\left. - \gamma^\nu D^{\mu,jk} \right) 
- m_{\Delta0} \delta^{jk}\left(  g^{\mu\nu} - \gamma^{\mu}\gamma^\nu \right) \nonumber \\
&&+ \frac{g_1}{2}\slashed{u}^{jk}\gamma_5g^{\mu\nu} \bigg\}
\xi^{\frac{3}{2}}_{kl}{\Psi}_\nu^l\ ,
\eea 
and
\bea\label{lag:piND}
{\cal L}^{(1)}_{\pi N\Delta}&=&h_A\,\bar{\Psi}_{\mu}^i\xi_{ij}^{\frac{3}{2}}  
\omega^{\mu,j}\Psi+h.c.\ ,
\eea
where $\Psi_\nu$ is the vector-spinor isovector-isospinor Rarita-Schwinger field of  the $\Delta$-resonance
with a bare mass $m_{\Delta 0}$ and $\xi^{\frac{3}{2}}_{ij} = \delta_{ij} - \tau_i \tau_j /3  $ are the isospin-$3/2$ projectors. Explicit expressions of $\xi^{\frac{3}{2}}_{kl}{\Psi}_\nu^l$ in terms of the physical $\Delta$ states can be found in Eq.~(3.8) of Ref.~\cite{Yao:2016vbz}. The covariant derivative acting on the $\Delta$ fields is defined by
\bea
{\cal D}_{\mu,ij}&=&\left(\partial_\mu\delta_{ij}-2i\epsilon_{ijk}\Gamma_{\mu,k}
+\delta_{ij}\Gamma_{\mu}\right)\ ,
\eea
with $\Gamma_{\mu,k}=\frac{1}{2}\langle\tau_k\Gamma_\mu\rangle$, the brackets $\langle ... \rangle $ denoting the trace in isospin space. Finally, $\omega^{\alpha,i}=\frac{1}{2}\,\langle\tau^i u^\alpha \rangle$. Lagrangians ${\cal L}^{(1)}_{\pi\Delta}$ and ${\cal L}^{(1)}_{\pi N\Delta}$  are consistent in the sense that they are invariant under the so-called point transformation~\cite{Moldauer:1956zz,Wies:2006rv}, so as to compensate the spurious unphysical components of the Rarita-Schwinger field. 

Next to leading order contributions to the isovector axial form factor arise also from 
\bea
\mathcal{L}_{\pi N}^{(3)}=\bar{\Psi}\big\{\frac{d_{16}}{2}\gamma^\mu\gamma_5\langle\chi_+\rangle u_\mu +\frac{d_{22}}{2}\gamma^\mu\gamma_5[D_\nu,F_{\mu\nu}^-]\big\}\Psi\ ,
\eea
where $\chi^+=u^\dagger\chi u^\dagger+u\chi^\dagger u$ with $\chi={\rm diag}(M_\pi^2,M_\pi^2)$ and $F_{\mu\nu}^-=[D_\mu,u_\nu]-[D_\nu,u_\mu]$. These counterterms absorb divergences stemming from the loops.

\subsection{Leading one-loop results\label{sec:loop}}
\begin{figure}[t]
\vspace{1.cm}
\begin{center}
\epsfig{file=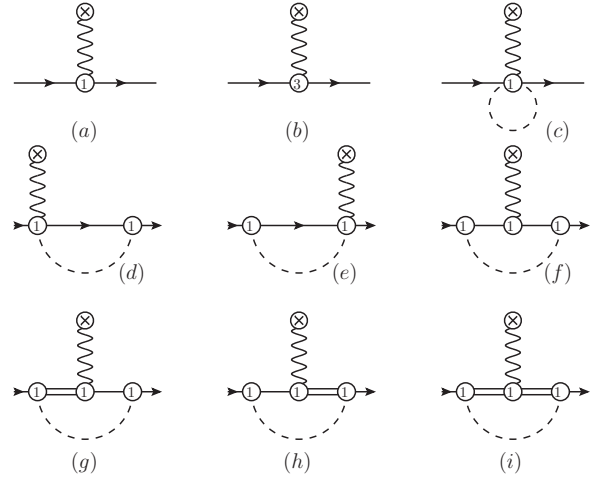,scale=0.56}
\caption{One-particle-irreducible Feynman diagrams contributing to $G_A(t)$ up to leading one-loop order. Dashed, solid and double lines represent pions, nucleons and $\Delta$ resonances, respectively. Numbers in the circles mark the chiral orders of the vertices.}
\label{fig:gA}
\end{center}
\end{figure}

The relevant Feynman diagrams contributing up to our accuracy, i.e., leading one-loop order, are shown in Fig.~\ref{fig:gA}. The $\Delta$-less diagrams~(a)-(f) have been calculated, e.g., in Ref.~\cite{Schindler:2006it}. We obtain the same results except for the sign of the term proportional to $d_{22}$ from diagram~(b). The calculations of diagrams~(g)-(i) are more complicated due to the complexity of the $\Delta$ propagator. Their contributions to the axial form factor are too lengthy to be shown explicitly.  The nucleon wave function renormalization constant $Z_N$ can be calculated from its corresponding self-energy, including $\Delta$ intermediate states~\cite{Alvarez-Ruso:2013fza,Yao:2016vbz}. The leading-order tree-level diagram (a), multiplied by $Z_N$, generates a loop-level contribution of $\mathcal{O}(p^3)$, which we denote $G_A^{(wf)}$. In summary, the unrenormalized leading one-loop axial form factor in BChPT reads
\bea
G_A(t)&=&g+4 d_{16} M_\pi^2 + d_{22} t + G_A^{(c)} +G_A^{(f)}\nonumber\\
&+&   2\,G_A^{(g)}+ G_A^{(i)}+G_A^{(wf)}\ ,
\eea
where $G_A^{(d)}=G_A^{(e)}=0$ and $G_A^{(g)}=G_A^{(h)}$ are implied.

The UV divergences stemming from loops are subtracted within the $\overline{\rm MS}-1$ (or $\widetilde{\rm MS}$) scheme. Specifically, the UV divergences are canceled by the infinite parts in the bare parameters  $g$, $d_{16}$ and $d_{22}$. The remaining UV-finite parts are denoted by $\bar{g}$, $\bar{d}_{16}$ and $\bar{d}_{22}$, respectively.

In the SSE counting, loops contribute at $\mathcal{O}(p^3)$. However, there are PCB terms due to the presence of internal matter fields, $N$ and $\Delta$, in the loops~\cite{Gasser:1987rb}. To restore the power counting, we adopt the EOMS scheme proposed in Refs.~\cite{Gegelia:1999gf,Fuchs:2003qc}. Here it means that an additional finite shift of $\bar{g}$ is carried out to cancel the PCB terms, finally leading to an EOMS-renormalized constant $\gA$, 
which corresponds to the axial coupling in the chiral limit (see below). Eventually, we get the renormalized axial form factor,
\bea\label{eq:GAq2ren}
G_A(t)&=&\gA+4 \bar{d}_{16} M_\pi^2 + \bar{d}_{22} t+\bar{G}_A^{(c)}+\bar{G}_A^{(f)}\nonumber\\
&+&   2\,\bar{G}_A^{(g)}+ \bar{G}_A^{(i)}+\bar{G}_A^{(wf)}\ ,
\eea
where the bar over each loop contribution indicates that both the UV divergences and the PCB terms have been subtracted. Note that both the nucleon and $\Delta$ bare masses, $m$ and $m_{\Delta 0}$ have been replaced by the physical ones, $m_N$ and $m_\Delta$, once the resulting differences are of higher orders. Furthermore, the $\Delta$ couplings $g_1$ and $h_A$ are also untouched by the renormalization procedure.

As stated in section~\ref{sec:definitions}, the axial charge is defined as the axial form factor at $t=0$, 
\bea\label{eq:gAren}
g_A=\gA+4\bar{d}_{16} \mpi^2+\frac{1}{16\pi^2F_\pi^2}g_A^{\rm loop}\ .
\eea
Moreover, in view of Eq.~(\ref{eq:rAdef}), the slope of the axial form factor at $t=0$ leads to the axial radius
\bea\label{eq:rAren}
\langle r_A^2\rangle =\frac{6}{\gA}\left[\bar{d}_{22}+\frac{1}{16\pi^2F_\pi^2}\langle {r}_A^2\rangle^{\rm loop}\right]\ .
\eea
The explicit expressions of the loop contributions $g_A^{\rm loop}$  and $\langle {r}_A^2\rangle^{\rm loop}$ are given in Eqs.~(\ref{eq:gAloop},\ref{eq:rAloop}). 

\section{Analysis of lattice QCD data\label{sec:lat}}

\subsection{Fitting procedure: $\Delta$-less vs $\Delta$-full\label{sec:fit}}

 \begin{figure}[t]
\vspace{1.cm}
\begin{center}
\epsfig{file=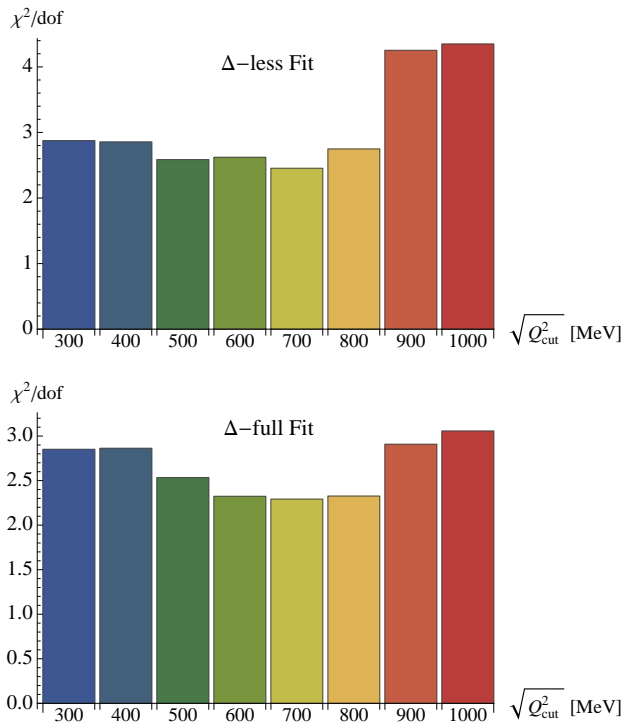,scale=0.56}
\caption{The values of $\chi^2/{\rm dof}$ for fits up to various $Q^2_{\rm cut}$. }
\label{fig:chisq}
\end{center}
\end{figure}

 \begin{figure*}[t]
\vspace{1.cm}
\begin{center}
\epsfig{file=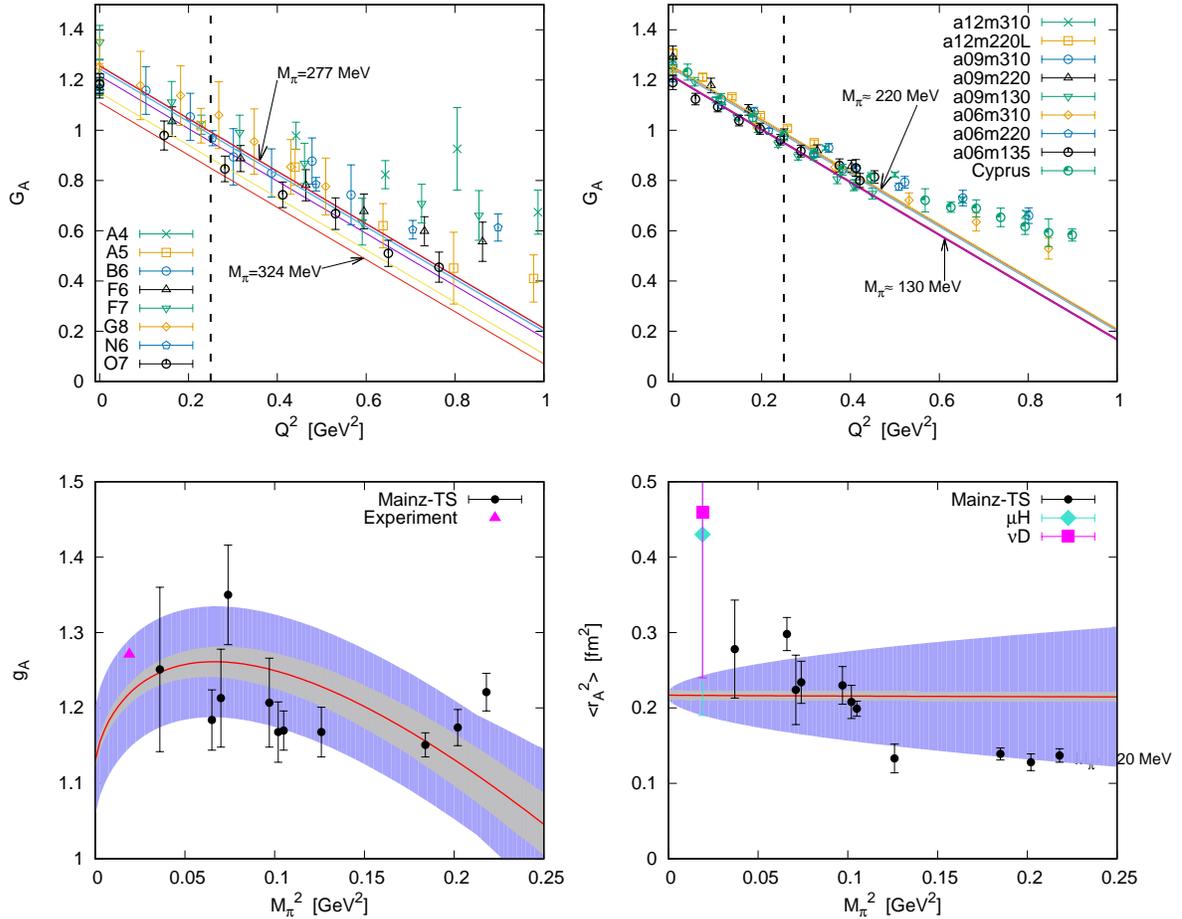,scale=1.0}
\caption{\label{fig:gAandrA.deltaless}
Results of ($\Delta$-less) Fit-I. Upper panels: fit results for the axial form factor. The dashed vertical line shows the value of $Q^2_{\rm cut}=0.25$~GeV$^2$: only points on the left have been used in the fit. The data ensembles A4-O7 (two-state method), a12m310-a06m135 and Cyprus (two-state method), are taken from Ref.~\cite{Capitani:2017qpc}, Ref.~\cite{Rajan:2017lxk} and Ref.~\cite{Alexandrou:2017hac}, respectively. In the top-left panel, the solid lines stand for the chiral representations of $G_A(Q^2)$ with various pion masses, ranging from 193~MeV to 364~MeV, corresponding to the different ensembles. In the top-right panel, pion masses are ranging from 130~MeV to 310~MeV. More specificaly, pion masses around $130$~MeV, around $220$~MeV and around $310$~MeV are used and hence the nine solid lines of $G_A(Q^2)$ overlap into very narrow bands. The pion masses corresponding to boundary lines are indicated explicitly.  
Left lower panel: prediction of the pion mass dependence of the axial charge. The black dots with error bars are the lattice determinations in Ref.~\cite{Capitani:2017qpc}, while the magenta triangle without error bar (the error is too tiny to be shown) stands for the precise experimental value~\cite{Patrignani:2016xqp}. Right lower panel:  prediction of the pion mass dependence of the axial radius. For comparison, the results obtained using the $z$-expansion are also shown as dots with error bars~\cite{Capitani:2017qpc}; the magenta square with error bars represents the recent extraction from neutrino quasi-elastic scattering data on deuterium ($\nu$D)~\cite{Meyer:2016oeg}; the turquoise diamond  with errors stands for the determination from the weak capture rate in muonic hydrogen ($\mu$H)~\cite{Hill:2017wgb}. The inner bands represent the statistical errors obtained by varying the LECs within their $1$-$\sigma$ uncertainties. The outer bands stand for the total errors where the theoretical uncertainties (see the text for details) are added to the statistical errors in quadrature .}
\end{center}
\end{figure*}

 \begin{figure*}[t]
\vspace{1.cm}
\begin{center}
\epsfig{file=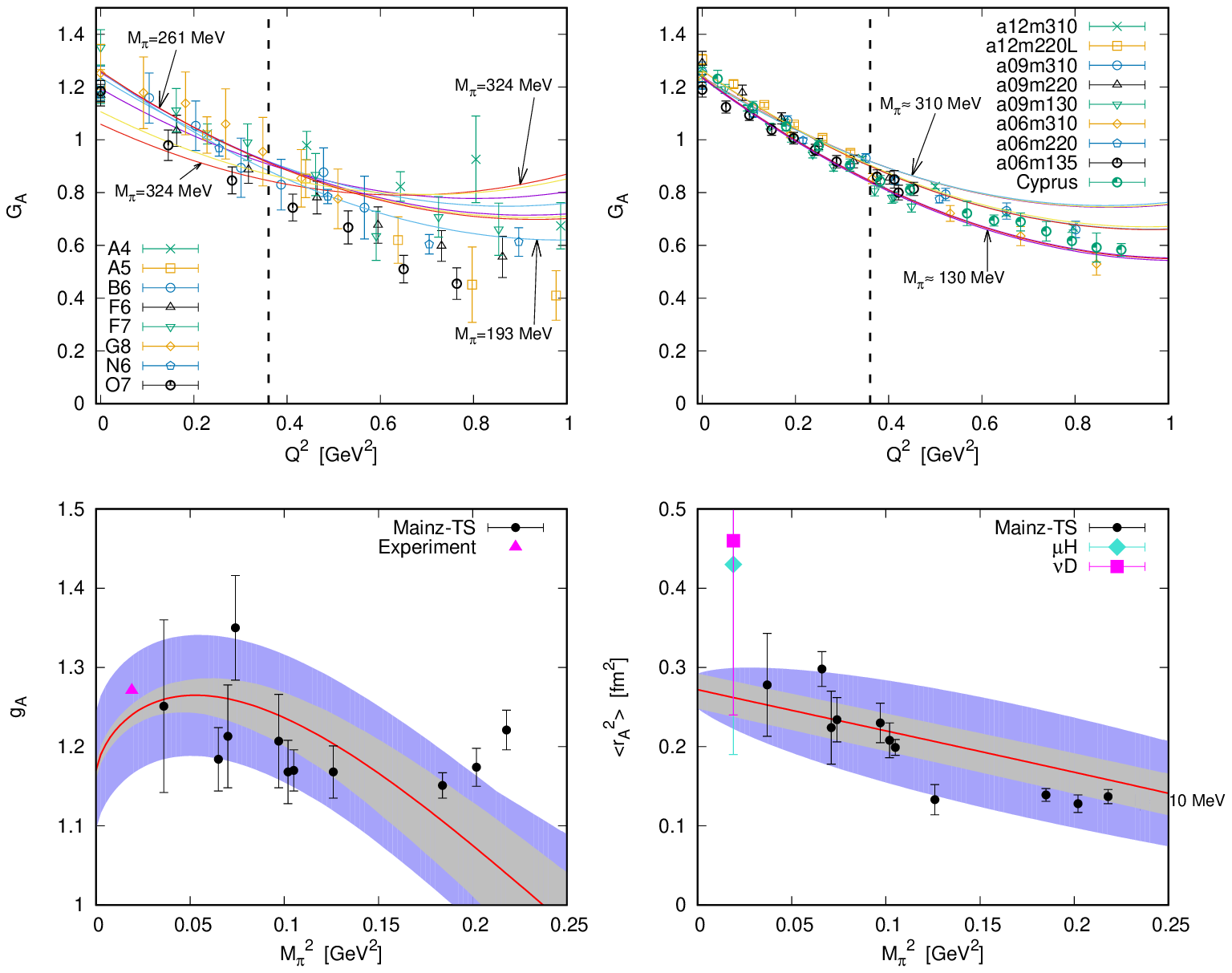,scale=1.0}
\caption{\label{fig:gAandrA}Results of $\Delta$-full fit and corresponding predictions. The description is the same as in Fig.~\ref{fig:gAandrA.deltaless} except that now $Q^2_{\rm cut}=0.36$~GeV$^{2}$. }
\end{center}
\end{figure*}

Axial form factor results by several groups are available:  RBC and UKQCD Collaborations~\cite{Yamazaki:2009zq}, LHPC~\cite{Bratt:2010jn}, ETM~\cite{Alexandrou:2010hf}, Ref.~\cite{Green:2017keo}, Ref.~\cite{Alexandrou:2017hac} (labeled by us as 'Cyprus'), Ref.~\cite{Capitani:2017qpc} (labeled as 'Mainz') and PNDME~\cite{Rajan:2017lxk}. In our fit procedure below, we include the latest lattice QCD data obtained using the two-state method by the Cyprus~\cite{Alexandrou:2017hac} with $M_\pi=130$~MeV, Mainz~\cite{Capitani:2017qpc}  collaborations with $M_\pi\in[193,473]$~MeV, as well as the data by PNDME~\cite{Rajan:2017lxk} collaboration with $M_\pi\sim [130,220,310]$~MeV. These lattice data have the advantage that systematical errors are better controlled by using improved techniques. In particular, the excited-state contamination is carefully considered. 

To assess the role of $\Delta$ degrees of freedom, we perform fits without and with $\Delta$ contributions, which are denoted as $\Delta$-less fit (Fit-I) and $\Delta$-full fit (Fit-II).  Due to the limited range of chiral perturbative calculations, only the data in the range $[0,Q^2_{\rm cut}]$ are taken into account, where $Q^2_{\rm cut}=0.25$~GeV$^2$ for the $\Delta$-less fit and $Q^2_{\rm cut}=0.36$~GeV$^2$ for the $\Delta$-full fit. The above two values are chosen such that plateau-like behaviors start to appear when further increasing $Q^2_{\rm cut}$, as one can see from Fig.~\ref{fig:chisq}, where values of $\chi^2/{\rm dof}$ ("dof" is an abbreviation for "degree of freedom") for fits up to various $Q^2_{\rm cut}$ are shown. The results of the $\Delta$-less fit are worse if the same $Q^2_{\rm cut}$ as that of the $\Delta$-full fit is used.
Furthermore, Mainz ensembles A3, E5 and N5 with $M_\pi\ge400$~MeV are excluded from our fits because such a pion mass is too large for the chiral extrapolation under our current accuracy. We have checked that their contribution to the total $\chi^2$ function is indeed large. 

 \begin{table}[ht]
\caption{Values and correlations of the LECs from Fit-I with $Q^2_{\rm cut}=0.25$~GeV$^2$. }\label{tab:fitI}
\vspace{-0.5cm}
\bea
\begin{array}{lr|ccc}
\hline\hline
&\text{Value}&\multicolumn{3}{c}{\text{Correlation matrix}}\\
\cline{3-5}
\chi^2/{\rm dof}&\frac{121.5}{50-5}~~~& \gA & \bar{d}_{16}  & \bar{d}_{22} \\
\hline
\gA&1.13(1)&1&-0.84&0.66\\
\bar{d}_{16}~[{\rm GeV}^{-2}]&-0.83(3)&&1&-0.24\\
\bar{d}_{22}~[{\rm GeV}^{-2}]&0.96(3)&&&1\\
\hline\hline
\end{array}\nonumber
\eea
\end{table}
 
 In our numerical computation, we employ the following values for the physical masses and the pion decay constant: $M_\pi^{\rm phy}=135$~MeV, $m_N=939$~MeV, $m_\Delta=1232$~MeV and $F_\pi=0.922$~MeV. The dimensional regularization scale is set equal to this nucleon mass.    
 In the chiral representation of $G_A$ in Eq.~(\ref{eq:GAq2ren}), there are five LECs: $\gA$, $\bar{d}_{16}$, $\bar{d}_{22}$, $g_1$ and $h_A$. Fit I is carried out by switching off the contributions from $\Delta$-resonance, i.e., setting $g_1=h_A=0$. The results for the fitted LECs are compiled in Table~\ref{tab:fitI}. The LECs values turn out to be of a natural size and the correlations are small. The corresponding plots for $G_A(Q^2)$ are shown in the upper panels of Fig.~\ref{fig:gAandrA.deltaless}. These $\Delta$-less chiral results for $G_A$ (solid lines in the figure) exhibit a linear dependence in $Q^2$. In the fitting range of $Q^2\in[0,0.25~{\rm GeV}^2]$ they are well compatible with the lattice QCD data, while above $Q^2\sim0.5~{\rm GeV}^2$ apparent discrepancies start to appear. 
 
  \begin{table}[ht]
\caption{Values and correlations of the LECs from Fit II with $Q^2_{\rm cut}=0.36$~GeV$^2$. The $\ast$ denotes an input value.}\label{tab:fit.3p}
\vspace{-0.5cm}
\bea
\begin{array}{lr|ccc}
\hline\hline
&\text{Value}&\multicolumn{3}{c}{\text{Correlation matrix}}\\
\cline{3-5}
\chi^2/{\rm dof}&\frac{146.4}{66-3}~~~& \gA & \bar{d}_{16}  & \bar{d}_{22} \\
\hline
\gA&1.17(1)&1&-0.66&0.66\\
\bar{d}_{16}~[{\rm GeV}^{-2}]&1.27(2)&&1&0.03\\
\bar{d}_{22}~[{\rm GeV}^{-2}]&5.20(2)&&&1\\
h_A&1.42^\ast&&\\
g_1&-1.21^\ast&&\\
\hline\hline
\end{array}\nonumber
\eea
\end{table}
 
 As for the $\Delta$-full fit, i.e. Fit-II, we first treat all the five LECs as free parameters. We find that the correlations among $\bar{d}_{16}$, $g_1$ and $h_A$ are quite large. Such large correlations lead to large errors. Besides, we have also checked that the five-parameter fit is very sensitive to the initial values of the fitting parameters. To tackle this issue, one has to fix at least two of them in the fit. Therefore, we improve our fit by using the central values, $h_A=1.42$ and $g_1=-1.21$, determined from $\pi N$ scattering in Ref.~\cite{Yao:2016vbz}, where the calculation was done up to the leading one-loop order in SSE scheme as well.  The errors of $h_A$ and $g_1$ are not taken into account in the fit but will be included in the error budget presented in the next subsection. The resulting best-fit parameters are shown in Table~\ref{tab:fit.3p}. Compared to Fit-I, the quality of Fit-II is better since the $\chi^2/{\rm dof}$ is smaller, in spite of the fact that the fit range is extended up to $Q^2=0.36~{\rm GeV}^2$ and the same number of fit parameters is used. Furthermore, the presence of $\Delta$ loops has a significant impact on the  $\bar{d}_{16}$ and $\bar{d}_{22}$ values. Plots of $G_A$ are shown in the upper panels of Fig.~\ref{fig:gAandrA}. The $Q^2$-dependence of $G_A$ is improved due to the inclusion of the loop contributions involving the $\Delta$-resonance. For the sake of completeness and to show the analytic behavior of our results, we have plotted the form factor beyond the fitting region and up to $Q^2 \sim 1$~GeV$^2$ both in Figs.~\ref{fig:gAandrA.deltaless} and \ref{fig:gAandrA}. As the applicability of BChPT breaks down at high $Q^2$, the (dis)agreement of our results with lattice data in this $Q^2$ region should be regarded as accidental.

\subsection{Extraction of the axial charge and radius\label{sec:gara}}

Based on the fitted values on Tables~\ref{tab:fitI} and \ref{tab:fit.3p}, we can extract the axial charge $g_A$ and squared radius $\langle r_A^2\rangle$ by using Eq.~(\ref{eq:gAren}) and Eq.~(\ref{eq:rAren}), respectively. As error budget, we take into account two kinds of uncertainties. On the one hand, the statistical errors are propagated from the fitted parameters from a Monte Carlo simulation considering the normal distributions of the parameters in Table~\ref{tab:fit.3p}. The error of $h_A=1.42(2)$ is also considered. As for $g_1$, it only appears in the loop contribution at next-to-next-to leading order in pion-nucleon scattering and hence can only be determined with a large error~\cite{Yao:2016vbz}. To avoid overestimating  statistical errors from $g_1$, we just use in the Monte Carlo simulation values which satisfy the $\Delta$-width constraint on $g_1$ and $h_A$, i.e., Eq.~(11) of Ref.~\cite{Gegelia:2016pjm}. Note that we demand the $\Delta$ width takes its Breit-Wigner value of $117\pm 3$~MeV, as quoted by PDG~\cite{Patrignani:2016xqp}. On the other hand, the theoretical error is estimated by truncation of the chiral series. We follow the method developed in Ref.~\cite{Epelbaum:2014efa}, where the chiral  theoretical uncertainty of a prediction for a quantity $O$ up to $\mathcal{O}(p^n)$ is assigned to
\bea\label{eq:error}
\delta\,O^{(n)}_{\rm theo.}&=&{\rm max}\big(| O^{n_{\rm LO}} | \mathcal{Q}^{n-n_{\rm LO}+1},
\{|O^{(k)}-O^{(j)}|\mathcal{Q}^{n-j}\}\big)\ ,\nonumber\\
&&n_{\rm LO}\leq j< k\leq n\, ,
\eea
with $n_{\rm LO}$ the leading chiral order. In our case we have $O\in\{g_A,\langle r_A^2\rangle\}$ and use $\mathcal{Q}=M_\pi/\Lambda_b$ with $\Lambda_b\sim4\pi F_\pi$ being the breakdown scale of the chiral expansion. 
Besides, the theoretical error in Eq.~(\ref{eq:error}) is required to be larger than the actual higher-order contribution,
\bea
\delta\,O^{(n)}_{\rm theo.}\geq {\rm max}\{|O^{(k)}-O^{(j)}|\}\ ,\qquad k\ge j \geq n\ .
\eea
For the purpose of this estimate, we calculate the diagrams of $O(p^4)$. There are only two diagrams, which have the same topologies as diagrams~(d) and (e) in Fig.~\ref{fig:gA} but the vertices containing the axial current are now of $O(p^2)$. The involved LECs are set to the values given by Fit~II(a)-$O(p^4)$ to pion-nucleon scattering data in Ref.~\cite{Chen:2012nx}. Moreover, the pion mass in their contribution is fixed to its physical value. Otherwise, the width of the theoretical error bands would increase extremely fast with the pion mass.

Eventually, the pion-mass dependences of $g_A$ and $\langle r_A^2\rangle$ are displayed in the lower panels of Figs.~\ref{fig:gAandrA.deltaless} and \ref{fig:gAandrA}, based on Fit-I and Fit-II, respectively.  The inner error bands represent the statistical errors. The outer bands correspond to the total errors where the theoretical and statistical errors are added in quadrature.  For $g_A$, the chiral predictions both in the $\Delta$-less and  $\Delta$-full cases are in good agreement with the lattice results in Ref.~\cite{Capitani:2017qpc} below the pion mass of $400$~MeV. Above it, some of the lattice data are out of the error bands of the axial charge. This is not surprising since we only fit the axial form factor to the data with $M_\pi < 400$~MeV. A clear improvement in the description of $\langle r_A^2\rangle$ with the inclusion of the explicit $\Delta$ contribution is apparent from the lower-right panels.

 \begin{figure}[t]
\vspace{1.cm}
\begin{center}
\epsfig{file=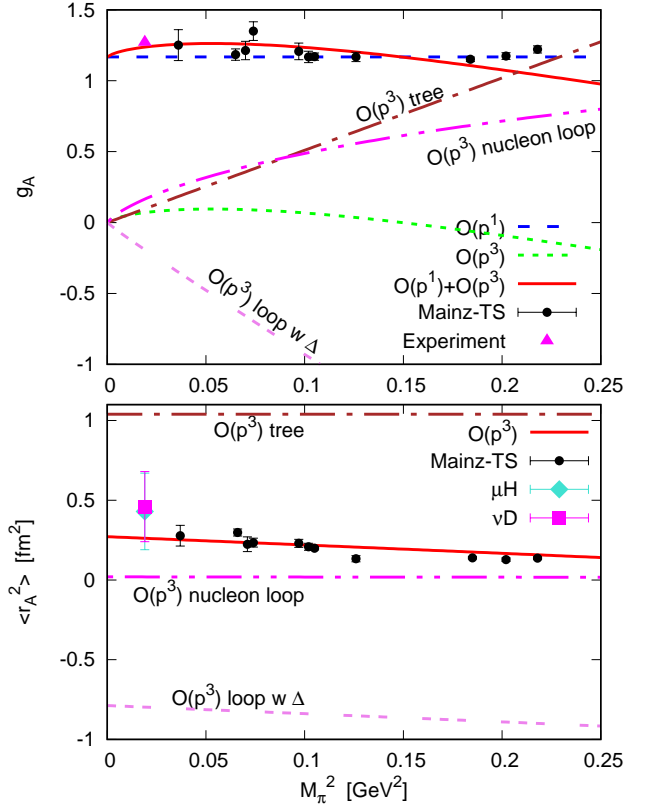,scale=1.1}
\caption{Convergence of axial charge and radius. Contributions from different types of Feynman diagrams using Fit-II parameters are displayed. As benchmark, we show the results from Ref.~\cite{Capitani:2017qpc}, which are denoted by the dots with error bars. The magenta triangle stands for the precise experimental value of $g_A$~\cite{Patrignani:2016xqp}. The magenta square and turquoise diamond, with error bars, represent the recent extractions from data for neutrino quasi-elastic scattering on deuterium ($\nu$D)~\cite{Meyer:2016oeg} and weak capture in muonic hydrogen ($\mu$H)~\cite{Hill:2017wgb}, respectively, using the $z$-expansion.}
\label{fig:gA.con}
\end{center}
\end{figure}

In Fig.~\ref{fig:gA.con}, we show the convergence properties of the nucleon axial charge and radius using Fit-II parameters. The respective  tree level, $\Delta$-less loop and full loop contributions at $\mathcal{O}(p^3)$ are displayed as well. We find that the axial charge converges rapidly and the loop terms, including $\Delta$, play a significant role in the convergence. Indeed, there exists a large cancellation between the $\Delta$-less and $\Delta$-loop (loops involving $\Delta$) contributions at $\mathcal{O}(p^3)$ level. As for the axial radius, the chiral series start to contribute at $\mathcal{O}(p^3)$, and we are not able to properly assess  its convergence within our current accuracy. Analogously to $g_A$, a cancellation takes place: the loop contribution involving internal $\Delta$ is negative while the other terms are positive.

 \begin{table}[ht]
\caption{Predictions for axial charge, radius and mass at physical pion mass. The numbers in brackets correspond to the total uncertainties obtained by adding
statistical and theoretical errors in quadrature.}\label{tab:prediction}
\vspace{-0.5cm}
\bea
\begin{array}{c|crr}
\hline\hline
&\text{Fit-I }~({\Delta}\text{-less})&\text{Fit-II }({\Delta}\text{-full})\\
\hline
g_A&1.215(72)&1.237(74)\\
\langle r_A^2\rangle~[\text{fm}^2]&0.217(26)&0.263(38)\\
M_A~[\text{GeV}]&1.47(8)&1.33(10)\\
\hline\hline
\end{array}\nonumber
\eea
\end{table}

At last, at the physical pion mass, we obtain $g_A$ and $\langle r_A^2\rangle$, with the corresponding axial mass $M_A$, shown in Table~\ref{tab:prediction}. 
Although lower, both the extracted values of $g_A$ are consistent with the experimental determination, $g_A=1.2723(23)$~\cite{Patrignani:2016xqp}, when the errors are taken into account. The agreement is improved after the $\Delta$ loops are taken into account. Regarding $\langle r_A^2\rangle$, the result based on Fit-II is in agreement both with the recent extractions from neutrino quasi-elastic scattering data on deuterium $\langle r_A^2\rangle = 0.46(22)$~fm$^2$~\cite{Meyer:2016oeg} and from the weak capture rate in muonic hydrogen $\langle r_A^2\rangle = 0.43(24)$~fm$^2$~\cite{Hill:2017wgb} using the $z$-expansion. This observation indicates that the inclusion of the explicit contribution $\Delta$-resonance improves the determination $\langle r_A^2\rangle$ significantly. On the other hand, when compared to earlier determinations~\cite{Liesenfeld:1999mv,Bernard:2001rs,Bodek:2007ym} using the dipole ansatz of Eq.~(\ref{eq:dipole}), which are consistent with the one of Ref.~\cite{Meyer:2016oeg} but with much smaller error bars, the present values for the axial radius are too low. Besides the possible error-bar underestimation of dipole fits~\cite{Meyer:2016oeg}, small axial radii could arise from the fact that $\langle r_A^2\rangle$ almost linearly depends on the pion mass squared (see lower-right panels of Figs.~\ref{fig:gAandrA.deltaless} and \ref{fig:gAandrA}), which is a typical behavior of the $\mathcal{O}(p^3)$ contribution. Therefore, to improve the chiral determination of $\langle r_A^2\rangle$, a quadratic (or higher power) term of $M_\pi^2$ from at least $\mathcal{O}(p^5)$ (including two-loop amplitudes) might be needed. On the lattice side, further studies of excited states and volume effects may be required~\cite{Alexandrou:2017hac}.

\section{Summary\label{sec:sum}}
We have calculated the nucleon axial form factor up to $\mathcal{O}(p^3)$ in a covariant baryon chiral perturbation theory with pion, nucleon and $\Delta(1232)$ as degrees of freedom. The axial form factor at leading one-loop order is renormalized by making use of the EOMS scheme, which restores the correct power counting while respecting the analytic structure of the amplitudes. 
The pion-mass and momentum-transfer dependences of the axial form factor are investigated by performing fits to recent lattice QCD data both without and with explicit $\Delta$ contribution. 
Based on the fitted values of the involved LECs, we have studied the pion mass dependence of the axial charge and radius. We find that the inclusion of $\Delta$ improves the chiral description of lattice QCD data significantly. Hence, we quote  $g_A=1.237(74)$ and  $\langle r_A^2\rangle =0.263(38)~{\rm fm}^2$ from the $\Delta$-full fit as our final results for the axial charge and axial radius squared at the physical pion mass. This determination of $g_A$ is in agreement with its experimental value within uncertainty. The value of $\langle r_A^2\rangle$ is consistent with a recent extraction from neutrino quasielastic scattering data on deuterium, given the large error bars. However, it is still small compared to earlier values extracted from experimental data. Apart from other aspects of the experiment-based determination of nucleon axial form factors and their errors, these discrepancies can stem from the systematical uncertainties of the lattice QCD data or the underestimated theoretical error of the chiral expansion. A more precise determination of $\langle r_A^2\rangle$ demands a chiral perturbative calculation at least at $\mathcal{O}(p^5)$ where actual chiral corrections are accounted for.

\acknowledgements
We would like to thank D.~Djukanovic, J.~Gegelia, R. Hill and A. Kronfeld for helpful comments on the manuscript. This research is supported by the Spanish Ministerio de Econom\'ia y Competitividad and the European Regional Development Fund, under contracts FIS2014-51948-C2-1-P, FIS2014-51948-C2-2-P, SEV-2014-0398 and by Generalitat Valenciana under contract PROMETEOII/2014/0068.

\appendix
\section{Explicit expressions for  $g_A$ and $\langle {r}_A^2\rangle$ \label{sec:ologa}}
The following abbreviations are used: $\Delta=\mn-\md$, $\Sigma=\mn+\md$\,. The one-point one-loop function is defined by 
\bea
A_0(M_a^2)=M_a^2\ln\frac{M_a^2}{\mu^2} ,
\eea
where $\mu$ is renormalization scale in dimensional regularization.
The scalar two-point integral has the following
analytical form
\bea
B_0(p^2,M_a^2,M_b^2) &=&
1-\ln\frac{M_b^2}{\mu^2}+
\frac{M_a^2-M_b^2+p^2}{2\,p^2}\ln\frac{M_b^2}{M_a^2} \nonumber\\
&&\hspace{-2.0cm}+
\frac{p^2-(M_a-M_b)^2}{p^2}\rho_{ab}(p^2)
\ln\frac{\rho_{ab}(p^2)-1}{\rho_{ab}(p^2)+1}\ ,
\eea
with
\bea
\rho_{ab}(p^2)\equiv\sqrt{\frac{p^2-(M_a+M_b)^2}{p^2-(M_a-M_b)^2}}.
\eea
Note that the notations for $A_0$ and $B_0$ functions are introduced in Ref.~\cite{Passarino:1978jh} and one can also use the numerical package LoopTools~\cite{Hahn:1998yk} to calculate $A_0$ and $B_0$ by applying the $\overline{\rm MS}-1$ subtraction scheme.

The explicit expression of the loop contribution to $g_A$ is given by
\begin{widetext}
\bea\label{eq:gAloop}
g_A^{\rm loop}&=&\bigg\{\frac{4 \mpi^2 \mn^2 \gA^3}{ \left(\mpi^2-4 \mn^2\right)}+\frac{h_A^2 \mpi^2 \left(189 \mpi^2-16 \mn (31 \mn+42 \md)\right) \gA}{324  \md^2}-\frac{5 \go h_A^2 \mpi^2}{972  \md^4} \bigg[58 \mpi^4
\no\\
&+&\left(6 \mn^2+142 \md \mn+107 \md^2\right) \mpi^2-2 \md \left(40 \mn^3+48 \md \mn^2+44 \md^2 \mn+\md^3\right)\bigg]\bigg\}
\no\\
&+&\bigg\{\gA+\frac{4 \left(\mpi^2-2 \mn^2\right) \gA^3}{ \left(\mpi^2-4 \mn^2\right)}+\frac{\gA h_A^2 }{54  \mn^2 \md^2} \bigg[23 \mpi^4-\left(93 \mn^2+20 \md \mn+46 \md^2\right) \mpi^2
\no\\
&+&(\mn+\md)^2 \left(41 \mn^2-26 \md \mn+23 \md^2\right)\bigg]+\frac{5 \go h_A^2 }{486  \mn^2 \md^4}\bigg[-2 \mpi^6+(32 \mn^2\no\\
&+&14 \md \mn-3 \md^2) \mpi^4+\left(20 \mn^4+66 \md \mn^3+45 \md^2 \mn^2+20 \md^3 \mn+12 \md^4\right) \mpi^2
\no\\
&-&(\mn+\md)^2 \left(2 \mn^4+10 \md \mn^3-9 \md^2 \mn^2+20 \md^3 \mn+7 \md^4\right)\bigg]\bigg\}\, A_0\left(\mpi^2\right)
\no\\
&+&\bigg\{\frac{4 \gA h_A^2 \mpi^2 \left(\mpi^2-\Sigma (2 \mn+\md)\right)}{27  \mn \Delta \md^2}-\frac{4 \gA^3 \mpi^2}{ \left(\mpi^2-4 \mn^2\right)}\bigg\}\,A_0\left(\mn^2\right)
\no\\
&+&\bigg\{\frac{5 \go h_A^2 \mpi^2}{486  \mn^2 \md^4} \bigg[2 \mpi^4+\left(-6 \mn^2-14 \md \mn+3 \md^2\right) \mpi^2+6 \mn^4-12 \md^4
\nonumber\\
&-&20 \mn \md^3+8 \mn^2 \md^2+28 \mn^3 \md\bigg]+\frac{\gA h_A^2 \mpi^2}{54  \mn^2 (\mn-\md) \md^2}\bigg[\mpi^2 (23 \md-31 \mn)-2 \Sigma\nonumber\\
&\times& \left(5 \mn^2-40 \md \mn+23 \md^2\right)\bigg]\bigg\} \,A_0\left(\md^2\right)
+\bigg\{\Delta \Sigma^3\bigg[\frac{\gA h_A^2  }{54  \mn^2 \md^2}\left(41 \mn^2-26 \md \mn+23 \md^2\right)\no\\
&-&\frac{5 \go h_A^2  }{486  \mn^2 \md^4}\left(2 \mn^4+10 \md \mn^3-9 \md^2 \mn^2+20 \md^3 \mn+7 \md^4\right)\bigg]\bigg\}\, B_0\left(\mn^2,0,\md^2\right)
\no\\
&+&\bigg\{\frac{\left(8 \mpi^2 \mn^2-3 \mpi^4\right) \gA^3}{ \left(\mpi^2-4 \mn^2\right)}-{2 \mpi^2 \gA}+\frac{4 h_A^2 \mpi^2  \gA}{27  \mn \Delta \md^2}(\mpi^4-\left(4 \mn^2+3 \md \mn+\md^2\right) \mpi^2
\nonumber\\
&+&16 \mn^2 \md^2)\bigg\} \,B_0\left(\mn^2,\mpi^2,\mn^2\right)
+\bigg\{\frac{5 \go h_A^2((\mn+\md)^2-\mpi^2)  }{486  \mn^2 \md^4}\bigg[-7 \md^6-20 \mn \md^5
\nonumber\\
&+&4 \left(3 \mpi^2+4 \mn^2\right) \md^4+
10 \mn \left(\mpi^2+\mn^2\right) \md^3+\left(-3 \mpi^4+14 \mn^2 \mpi^2-11 \mn^4\right) \md^2
\no\\
&+&10 \mn \left(\mpi^2-\mn^2\right)^2 \md-2 \left(\mpi^2-\mn^2\right)^3\bigg]
-\frac{\gA h_A^2(\Sigma^2-\mpi^2)}{54  \mn^2 \Delta \md^2} 
\bigg[(23 \md-31 \mn) \mpi^4
\nonumber\\
&+&\left(-10 \mn^3+8 \md \mn^2+80 \md^2 \mn-46 \md^3\right) \mpi^2
\no\\
&+&\Delta^2 \Sigma \left(41 \mn^2-26 \md \mn+23 \md^2\right)\bigg]\bigg\} \,B_0\left(\mn^2,\mpi^2,\md^2\right)\ .
\eea
The explicit expression of the loop contribution to $\langle {r}_A^2\rangle$ reads
\bea\label{eq:rAloop}
\langle {r}_A^2\rangle^{\rm loop}&=&
\frac{-12 \gA^3
   \mpi^2 \mn^2}{ \left(\mpi^2-4
   \mn^2\right)^2}+\frac{ 4 \gA h_A^2 \mpi^2}{15  \mn
   \md^2 \Delta^2 \Sigma}\bigg
   [8 \mpi^4-2
   \mpi^2 \left(13 \mn^2+16 \mn \md+5
   \md^2\right)
   \nonumber\\
  & +&9 \mn^4+12 \mn^3 \md+48 \mn^2
   \md^2+28 \mn \md^3-\md^4\bigg]
   +
   \frac{5 \go h_A^2 \mpi^2}{81  \mn^2 \md^4
   \left(\Delta^2-\mpi^2\right)}
    \bigg[3 \mpi^6
    \nonumber\\
    &-&3 \mpi^4 \left(29
   \mn^2+5 \mn \md+\md^2\right)+\mpi^2 \left(132
   \mn^4-351 \mn^3 \md-274 \mn^2 \md^2+3 \mn
   \md^3-3 \md^4\right)
   \nonumber\\
   &-&(\mn-\md) \left(48
   \mn^5-252 \mn^4 \md-112 \mn^3 \md^2+514
   \mn^2 \md^3+15 \mn \md^4+3
   \md^5\right)\bigg]
   \nonumber\\
   &+&\bigg\{
      \frac{6 \gA^3
   \mpi^2 \left(\mpi^2-5 \mn^2\right)}{ \mn^2
   \left(\mpi^2-4 \mn^2\right)^2}
   +\frac{8 \gA h_A^2 }{15 
   \mn^3 \md^2 \Delta^2 \Sigma}\bigg[
 4
   \mpi^6-8 \mpi^4 \left(2 \mn^2+2 \mn
   \md+\md^2\right)
   \nonumber\\
   &+&\mpi^2 \left(15 \mn^4+18 \mn^3
   \md+30 \mn^2 \md^2+26 \mn \md^3+7
   \md^4\right)-\left(\Delta\Sigma\right)^2 \left(3
   \mn^2+7 \mn \md+3 \md^2\right)
   \bigg]
   \nonumber\\
   &+&
   \frac{10 \go h_A^2 }{27  \mn^4 \md^4
   \left(\Delta^2-\mpi^2\right)}
   \bigg[
   \mpi^8-\mpi^6 \left(4 \mn^2+5
   \mn \md+3 \md^2\right)
   +\mpi^4 (14 \mn^4+8
   \mn^3 \md
   \nonumber\\
   &+&\mn^2 \md^2+11 \mn \md^3+3
   \md^4)-\mpi^2 (11 \mn^6-32 \mn^5
   \md-17 \mn^4 \md^2+35 \mn^3 \md^3+7 \mn^2
   \md^4
   \nonumber\\
   &+&7 \mn \md^5+\md^6)-\mn \md
   \left(\mn^2-\md^2\right)^2 \left(17 \mn^2-10 \mn
   \md-\md^2\right)
   \bigg]
   \bigg\}A_0(\mpi^2)
   \nonumber\\
&+& \bigg\{
 -\frac{6 \gA^3 \mpi^2 \left(\mpi^2-6 \mn^2\right)}{
   \mn^2 \left(\mpi^2-4 \mn^2\right)^2}
   -\frac{8 \gA h_A^2
   \mpi^2 }{15  \mn^3 \md^2
   \Delta^3 \Sigma^2}
   \bigg[
   2 \mpi^6-\mpi^4 \left(9 \mn^2+8 \mn
   \md+7 \md^2\right)
   \nonumber\\
    &+&\mpi^2 \left(7 \mn^4+13 \mn^3
   \md+32 \mn^2 \md^2+25 \mn \md^3+5
   \md^4\right)-\mn (3 \mn^5+20 \mn^4 \md-15
   \mn^3 \md^2
   \nonumber\\
   &-&20 \mn^2 \md^3+40 \mn
   \md^4+20 \md^5)
   \bigg]
 \bigg\} A_0(\mn^2)+\bigg\{
 \frac{10 \go h_A^2 \mpi^2}{27  \mn^4 \md^4
   (\mpi^2-\Delta^2) }
   \bigg[
 3 \md^4 \left(\mpi^2+2
   \mn^2\right)
   \nonumber\\
   &+&\mn \md^3 \left(11 \mpi^2-26
   \mn^2\right)+\left(\mpi^2-\mn^2\right)^3-\md^2 \left(3
   \mpi^4+\mpi^2 \mn^2-26 \mn^4\right)+\mn \md
   (-5 \mpi^4
   \nonumber\\ 
   &+&3 \mpi^2 \mn^2+2 \mn^4)-6
   \mn \md^5-\md^6
    \bigg] 
   +\frac{8
   \gA h_A^2 \mpi^2 }{15  \mn^3
   \md^2 \Delta^3 \Sigma^2}
   \bigg[
   2 \mpi^6-\mpi^4 (13
   \mn^2+8 \mn \md
   \nonumber\\
   &+&3 \md^2)+\mpi^2 \left(17
   \mn^4+29 \mn^3 \md+30 \mn^2 \md^2+9 \mn
   \md^3-3 \md^4\right)-3 \mn^6-14 \mn^5 \md
   \nonumber\\
   &-&6
   \mn^4 \md^2-12 \mn^3 \md^3-26 \mn^2
   \md^4+6 \mn \md^5+7 \md^6
   \bigg]
 \bigg\}A_0(\md^2)-
 \bigg\{
 \frac{10 \go h_A^2 \Sigma^3}{27  \mn^3
   \md^3}\bigg[
    17 \mn^3
    \nonumber\\
    &-&27 \mn^2 \md+9 \mn
   \md^2+\md^3
   \bigg]
   +\frac{8 \gA h_A^2 \Delta
   \Sigma^2}{15  \mn^3 \md^2}\bigg[ 3
   \mn^2+7 \mn \md+3 \md^2\bigg]
 \bigg\}B_0(\mn^2,0,\md^2)
 \nonumber\\
 &+&
 \bigg\{
 -\frac{6 \gA^3 \mpi^2 \left(\mpi^4-6 \mpi^2 \mn^2+6
   \mn^4\right)}{ \mn^2 \left(\mpi^2-4
   \mn^2\right)^2}
   -\frac{8 \gA h_A^2 \mpi^2}{15  \mn^3 \md^2
   \Delta^3 \Sigma^2}
   \bigg[
   2
   \mpi^8-\mpi^6 (13 \mn^2+8 \mn \md
   \nonumber\\
   &&+7
   \md^2)+\mpi^4 \left(23 \mn^4+29 \mn^3
   \md+46 \mn^2 \md^2+25 \mn \md^3+5
   \md^4\right)-2 \mpi^2 \mn (19 \mn^4
   \md
   \nonumber\\
   &+&6 \mn^5+21 \mn^3 \md^2+15 \mn^2 \md^3+25 \mn
   \md^4+10 \md^5)+20 \mn^3 \md
   \left(\Delta\Sigma \right)^2
   \bigg]
 \bigg\}
 B_0(\mn^2,\mpi^2,\mn^2)
 \nonumber\\
 &+&\bigg\{
 \frac{8 \gA h_A^2 \left(\Delta^2-\mpi^2\right)
   \left(\mpi^2-\Sigma^2\right)^2 }{15  \mn^3
   \md^2 \Delta^3 \Sigma^2}
   \bigg[
   -2
   \mpi^4+\mpi^2 \left(9 \mn^2+4 \mn
   \md-\md^2\right)
   \nonumber\\
   &+&\Delta^2 \left(3 \mn^2+7
   \mn \md+3 \md^2\right)
   \bigg]
   -\frac{10
   \go h_A^2 }{27  \mn^4 \md^4
   (\mpi^2-\Delta^2) }
   \bigg[
   -\mpi^{10}+\mpi^8 (4 \mn^2
   \nonumber\\
   &+&5
   \mn \md+4 \md^2)-2 \mpi^6 \left(3 \mn^4+4
   \mn^3 \md+2 \mn^2 \md^2+8 \mn \md^3+3
   \md^4\right)+2 \mpi^4 (2 \mn^6
   \nonumber\\
   &+&9 \mn^5
   \md+13 \mn^4 \md^2+19 \mn^3 \md^3+3 \mn^2
   \md^4+9 \mn \md^5+2 \md^6)-\mpi^2
   (\Delta\Sigma)  (\mn^6
   \nonumber\\
   &+&32 \mn^5
   \md+17 \mn^4 \md^2-18 \mn^3 \md^3-17
   \mn^2 \md^4-8 \mn \md^5-\md^6)+\mn
   \md \left(\Delta\Sigma\right)^3 (17 \mn^2
   \nonumber\\
   &-&10
   \mn \md-\md^2)
   \bigg]
 \bigg\} B_0(\mn^2,\mpi^2,\md^2)\ .
\eea
\end{widetext}

\end{document}